\documentclass[useAMS,usenatbib]{mn2e}

\input epsf

\newcommand{\xmm} {{\sl XMM-Newton}}
\newcommand{\chandra} {{\sl Chandra}}
\newcommand{\rosat} {{\sl ROSAT}}
\newcommand{\cl} { Abell~2255}

\title[Abell 2255]{\xmm\ observations of Abell~2255 : \\ a test case of a
merger after `core-crossing'.} \author[I. Sakelliou,
T.J. Ponman]{Irini Sakelliou$^{1,2}$, Trevor J. Ponman$^{1}$,\\
$^{1}$School of Physics \& Astronomy, University of Birmingham,
Edgbaston, Birmingham B15 2TT \\ $^{2}$Max-Planck-Institute f\"{u}r
Astronomie, K\"{o}nigstuhl, 17, D-69117, Heidelberg, Germany \\ }

\begin{document}

\pagerange{\pageref{firstpage}--\pageref{lastpage}} \pubyear{2005}

\maketitle

\label{firstpage}

\begin{abstract}
It has been known that \cl\ is not a relaxed cluster, but it is
undergoing a merger. Here, we report on the analysis of the \xmm\
observations of this cluster.  The X-ray data give us the opportunity
to reveal the complexity of the cluster, especially its temperature
distribution.  The integrated spectrum is well fitted by a single
temperature thermal model, indicating a mean temperature of
$\sim$7~keV. However, the cluster is not isothermal at this
temperature: its eastern regions are significantly cooler, at
$\sim$5.5~keV, whilst towards the West the temperature reaches
$\sim$8.5~keV.

These temperature asymmetries can be explained if \cl\ has been
assembled recently by the merging of smaller subunits. It is now in
the phase after the cores of these subunits have collided (the
`core-crossing' phase) some 0.1-0.2~Gyr ago. A comparison with
numerical simulations suggests that it will settle down into a single
relaxed cluster in $\sim$(2-3)~Gyr.
\end{abstract}


\section{Introduction}

After the first X-ray observations, it became apparent that \cl\ is a
cluster that is currently under formation, growing by the accumulation
of smaller subunits. The \rosat\ images showed that the cluster is
elongated along the East-West direction, and that the centroid of the
X-ray emission does not coincide with any large cluster galaxy (Burns
et all. 1995, Feretti et al. 1997). The spectral analysis of Davis \&
White (1998) found significant temperature structure in its
intracluster medium (ICM), which they also attributed to a recent
merger event. Abell~2255 was observed by \chandra\ with the ACIS-I
detector for a total of 39~ksec. The \chandra\ data set, was used by
Davis, Miller, \& Mushotzky (2003), to investigate the X-ray
properties of the cluster galaxies. It has also been observed by \xmm\,
and Fig.~1(a) shows the \xmm\ mosaic. In Fig~1(b) we present an
overlay of the X-ray contours onto an optical image of the central
regions of \cl\. The X-ray contours are obtained from the \xmm\
observations that will be presented in the subsequent sections of this
paper.

Optically, as can be seen in Fig.~1(b), a very intriguing property of
\cl\ is that the brightest galaxies are arranged in a chain, whose
orientation coincides with the major axis of the elliptical X-ray
emission. The cluster has an unusually high velocity dispersion of
$\sim$1200~${\rm km \ s^{-1}}$, and the two brightest galaxies
[galaxies A and B in Fig.~1(b)] are separated by $\sim$2600~${\rm km \
s^{-1}}$ (Burns et al. 1995). Performing deep multicolour photometry
in a large field around the cluster, Yuan, Zhou, \& Jjang (2003)
showed that at radii $>$(10-15)~arcmin there are small groups of
galaxies, that appear to rotate around the central core of Abell~2255.

\begin{table*}
\caption{Pointing Information}\label{obs_info}
\begin{center}
 \begin{tabular}{ccccccccc}   \hline \hline

   (I)          &
(II)            &
(III)           &
(IV)            &
(V)             &
(VI)            &
(VII)
\\

Rev             &
Obs             &
$\alpha$~(2000)         &
$\delta$~(2000)         &
Instr.                  &
$Exp$           &
$Exp_{corr}$            &

\\
                &
                &
                &
                &
                &
ksec            &
ksec            &

\\
\hline

525            &
0112260501      &
17 12 59.920        &
+64 03 25.00        &
MOS1                    &
8.558  &
4.307  &
\\

        &
        &
        &
        &
MOS2    &
8.645  &
4.466  &
\\

        &
        &
        &
        &
PN      &
4.326   &
1.665   &
\\

548            &
0112260801      &
17 12 58.460                &
+64 04 49.20              &
MOS1                            &
16.468          &
10.810          &
\\

        &
        &
        &
        &
MOS2    &
16.504  &
10.657  &
\\

        &
        &
        &
        &
PN      &
11.918  &
4.068  &
\\

\hline

\end{tabular}
\vspace{0.2cm}
\begin{minipage}{16cm}
\small NOTES : (I)-revolution number; (II)-observation number;
(III)-pointing $Right Ascension$; (IV)-pointing $Declination$; (V)
EPIC Instrument; (VI) Exposure time (live time for the central CCD);
(VII) Reduced Exposure time, after the subtraction of the bright
background flares (see text for more details).
\end{minipage}
\end{center}
\end{table*}

In the radio \cl\ contains a central radio halo (see, for example,
Giovannini, Tordi \& Feretti 1999, and references there-in), and a
number of tailed radio galaxies. More recently, Govoni et al. (2005)
presented a high sensitivity radio image of \cl\ which reveals the
detailed structure of its radio halo and a possible radio relic. Radio
halos are rare radio sources, and they have been found in the inner, 1
Mpc of X-ray bright and hot clusters [see Giovannini \& Feretti (2000)
for some recent examples]. They locate the site of relativistic
electrons and magnetic fields in clusters. The presence of such a halo
in Coma, in conjunction with its X-ray morphology, motivated the
suggestion by Burns et al. (1994) that radio halos are fuelled by
cluster collisions, and are associated with clusters undergoing
disturbances from recent or on-going merging events.  Since the first
discoveries, significant advances have been made in understanding
their origin (e.g. Buote 2001).

Thus, the X-ray, optical and radio data provide evidence that \cl\ is
currently active, and other surrounding structures might be
interacting with it. This impression is not very surprising if one
thinks that \cl\ is a member of the rich North Ecliptic Pole
supercluster, that contains at least 21 galaxy clusters, as was
revealed by the analysis of the \rosat\ All-Sky Survey data by Mullis
et al. (2001).

We have observed \cl\ with \xmm\ in order to uncover its dynamical
state, decide on its past history and future evolution, and derive
vital information that would help us to test the results of the
numerical simulations of merging clusters. In this paper, we present
the analysis and results of the \xmm\ observations: the observations
are described in Section~2; Sections 3 and 4 are devoted to the
presentation of the X-ray properties of the cluster, as found from the
\xmm\ data analysis; in Section~5 we compare the \xmm\ and \rosat\
results; finally, in Section~6 we discuss a possible dynamical
scenario that can describe well the data.

The redshift of \cl\ is z=0.0806 (NED), and throughout this paper we
use $H_0=71~{\rm km \ s^{-1} \ Mpc^{-1}}$, $\Omega_{\rm M}$=0.3, and
$\Omega_{\rm \Lambda}$=0.7, giving a scale of 1.499 kpc/". The
Galactic hydrogen column for the direction of \cl\ is $N_{\rm H,G} =
2.6\times 10^{20}\ {\rm cm^{2}}$.

\section{\xmm\ observations}

\cl\ was observed by \xmm\ for a total of $\sim$25~ksec. The observation
was split into two: the first part was performed on the 22$^{nd}$
October 2002 (revolution=525), while the second one on the 7$^{th}$
December 2002 (revolution=548).  Information about these two pointings
is gathered in Table~\ref{obs_info}.  During both observations the
EPIC instruments were operating in the PrimeFullWindow (for MOS1 and
MOS2), and PrimeFullWindowExtended (PN), and the thin filter was used
for all imaging detectors.

\begin{figure*}
\begin{center} 
\leavevmode 
\epsfxsize 0.46\hsize
\epsffile{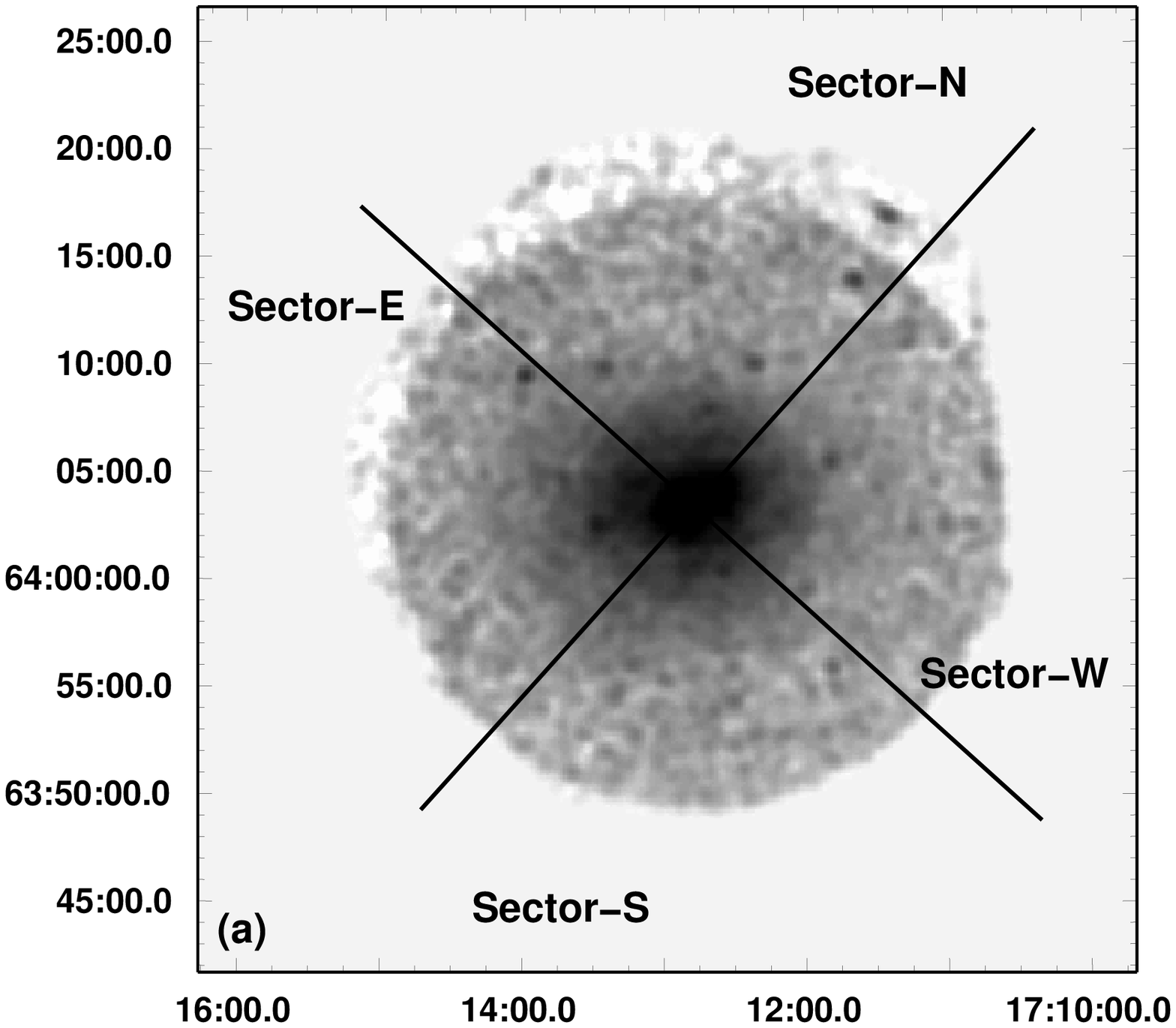}
\epsfxsize 0.45\hsize
\epsffile{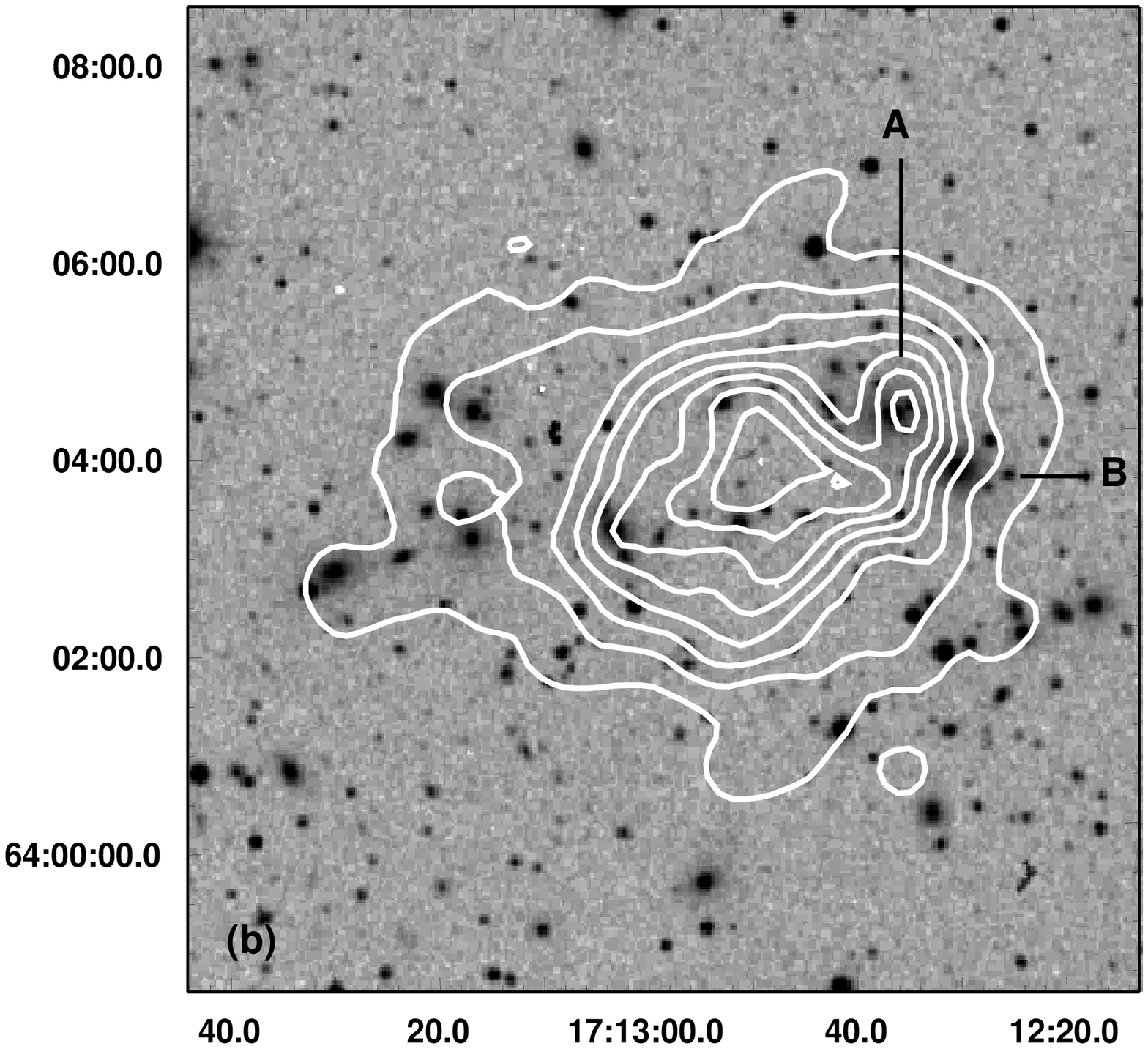}
\caption{(a) The \xmm\ mosaic of \cl\. Images from all the instruments
  and both observations in the (0.5-10.0)~keV energy range have been
  co-added. (b) DSS image of the central region of the cluster
  overlaid by the X-ray contours of the image shown in (a). The
  contour levels are linearly spaced from 1.4 to 6.3$\times 10^{-3} \
  {\rm cnt \ s^{-1} \ pix^{-2}}$, where each image pixel is
  8~arcsec.}\label{images}
\end{center} 
\end{figure*}

\subsection{Data Reduction}

The raw data from the EPIC instruments were processed with SAS
v5.4.1. {\sc emchain} and {\sc epchain} were used to obtain the
calibrated event lists for the MOS and PN instruments
respectively. During the processing the parameter {\it withbadpixfind}
was switched on, so that bad pixels that had not been recorded in the
calibration files were found and subsequently removed. After the
initial processing we confirmed that new bad pixels were found. The
calibrated events were filtered for {\sc flag}s, using the \xmm\ {\sc
flag}s {\sc $\#$xmmea\_em} and {\sc $\#$xmmea\_ep} for the two MOS and
the PN detectors respectively. Restrictions on the {\sc pattern} were
also applied: we kept only events with {\sc pattern}$<$12 for the MOS
cameras, and $<4$ for the PN.  We also cleaned the event lists for
periods of high background. This cleaning process reduced the exposure
times to those presented in column (VII) of Table~1.

\subsection{Background Treatment}

Background data were generated from the `blank-sky' event lists
(D. Lumb's background files; Lumb 2002). The coordinate frames of
these fields were converted to the corresponding frames for each \xmm\
pointing of the \cl\ observations. The background event lists were
filtered for {\sc pattern} and {\sc flag} in the same manner as was
done for the data.  Periods of high background levels that were still
present in D.Lumb's background files, were removed by applying a 3-$\sigma$
cut-off. Subsequently, the background events were scaled to match the
background levels of each instrument and observation by scaling the
out-of-field events as in Pratt et al. (2002). The scaling factors we
found for MOS1, MOS2 are 1.25 and 1.15 respectively for both
observations, consistent with previous findings. However, for the PN
detector we found scaling factors that are larger than expected (2.2
for the 0112260801 and 3.5 for the 0112260501 observation).

Given the uncertainty of the scaling factors for the PN detector, we
used the blank-sky background files only for the purpose of the spatial
analysis. The spectral analysis (see Section~\ref{spectral})
concentrates on the properties of the inner region of the cluster
($<$6~arcmin) leaving enough room for the safe use of local
background, as will be demonstrated in Section~\ref{spectral}. As a
test and to support our choice of the background files used for the
spectral analysis, we performed the same fitting procedures as the
ones presented in Section~4 adopting the blank-sky scaled files as the
background spectra. We found that the derived temperatures are
systematically higher than the ones derived with the local background
files by $\sim$(0.5-1.0)~keV, but they always show the same trends.

\section{Spatial Analysis}

For the purpose of any subsequent spatial analysis, firstly we created
background-subtracted and exposure-corrected images for both
observations and each EPIC instrument in the (0.5-10.0)~keV energy
range. The background and exposure correction was performed as 
in Sakelliou \& Ponman (2004). 

The background-subtracted and exposure-corrected images from each
camera and observation were added with the SAS task {\sc emosaic},
after scaling down the PN images to match the efficiency of the MOS
detectors. The final mosaic was smoothed with a Gaussian kernel of
$\sigma$=16~arcsec. Fig.~\ref{images}(a) presents this smoothed
mosaic. The same figure shows the orientation and boundaries of the
four sectors, that were used for the analysis in
Sections~\ref{spatial_sectors} and \ref{spectral_sectors}.  In
Fig.~\ref{images}(b) we overlay the X-ray contours onto a DSS image of
the central cluster region.

In these images, the cluster appears elongated along the East-West
direction, and the most sever disruptions are encountered within the
central $\sim$(8-10)~arcmin. As noted before, the peak of the X-ray
emission does not appear to coincide with any cluster galaxy.

In the following sections, we demonstrate how the background-subtracted
and exposure-corrected images were used to model the over-all light
distribution (Section~\ref{2d}), and in Section~\ref{spatial_sectors}
to investigate its azimuthal variations.

\subsection{2-dimensional analysis}\label{2d}

\begin{figure}
\begin{center} 
\leavevmode 
\epsfxsize 1.0\hsize
\epsffile{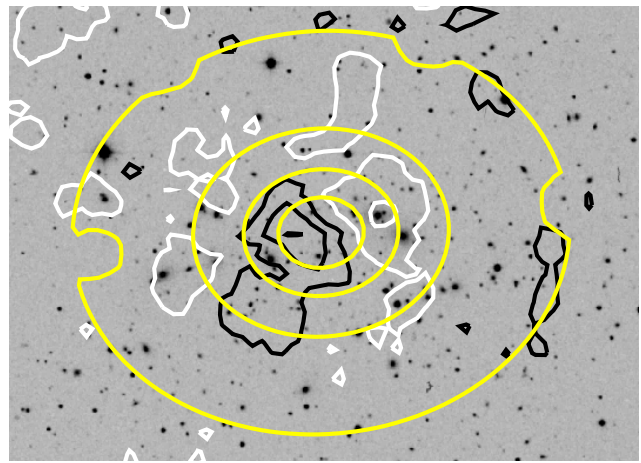}
\caption{Optical image of the central $\sim(17.3 \times
  12.4)$~arcmin$^2$ of \cl. The image is overlaid by the contours of
  the residuals, and the best fitting model
    derived from the  2-dimensional fit to the X-ray image (see 
    Section\,3.1). Positive residuals are shown as white lines,
    negative as black, and the best fitting
     model with the  grey ellipses.}\label{2D}
\end{center} 
\end{figure}

\begin{figure*}
\begin{center} 
\leavevmode 
\epsfxsize 0.48\hsize
\epsffile{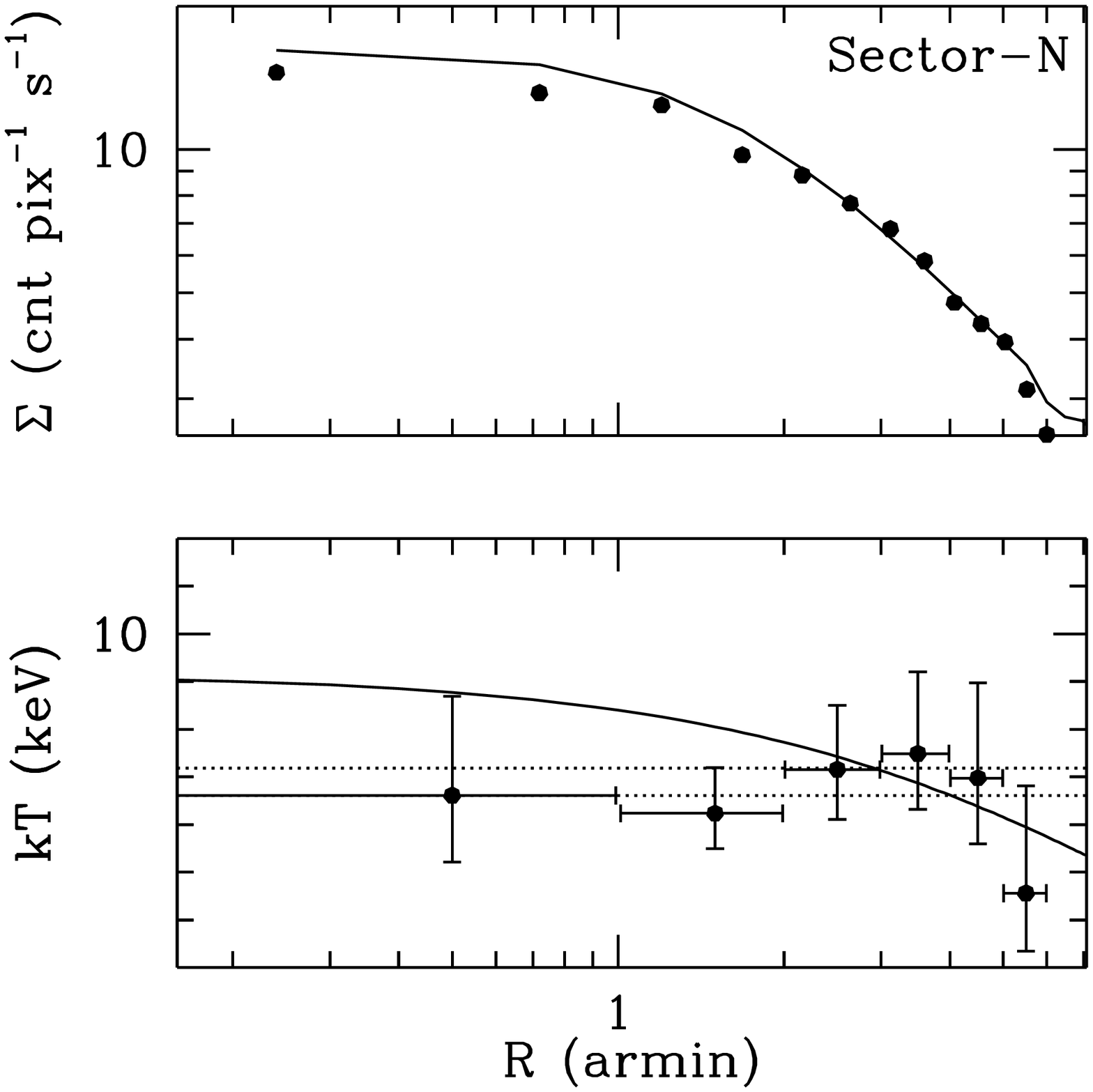}
\leavevmode 
\epsfxsize 0.48\hsize
\epsffile{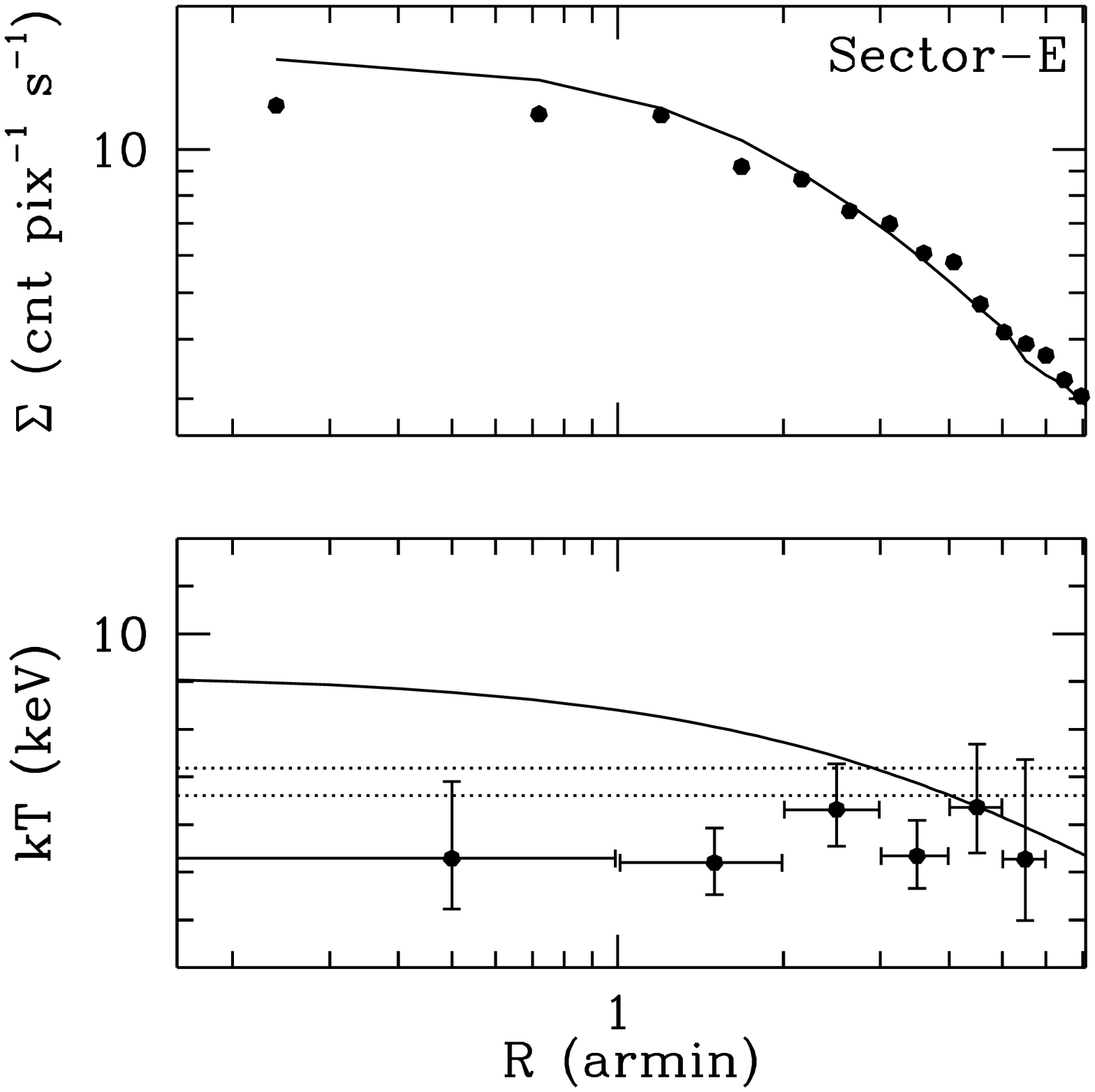}
\end{center}
\begin{center} 
\leavevmode 
\epsfxsize 0.48\hsize
\epsffile{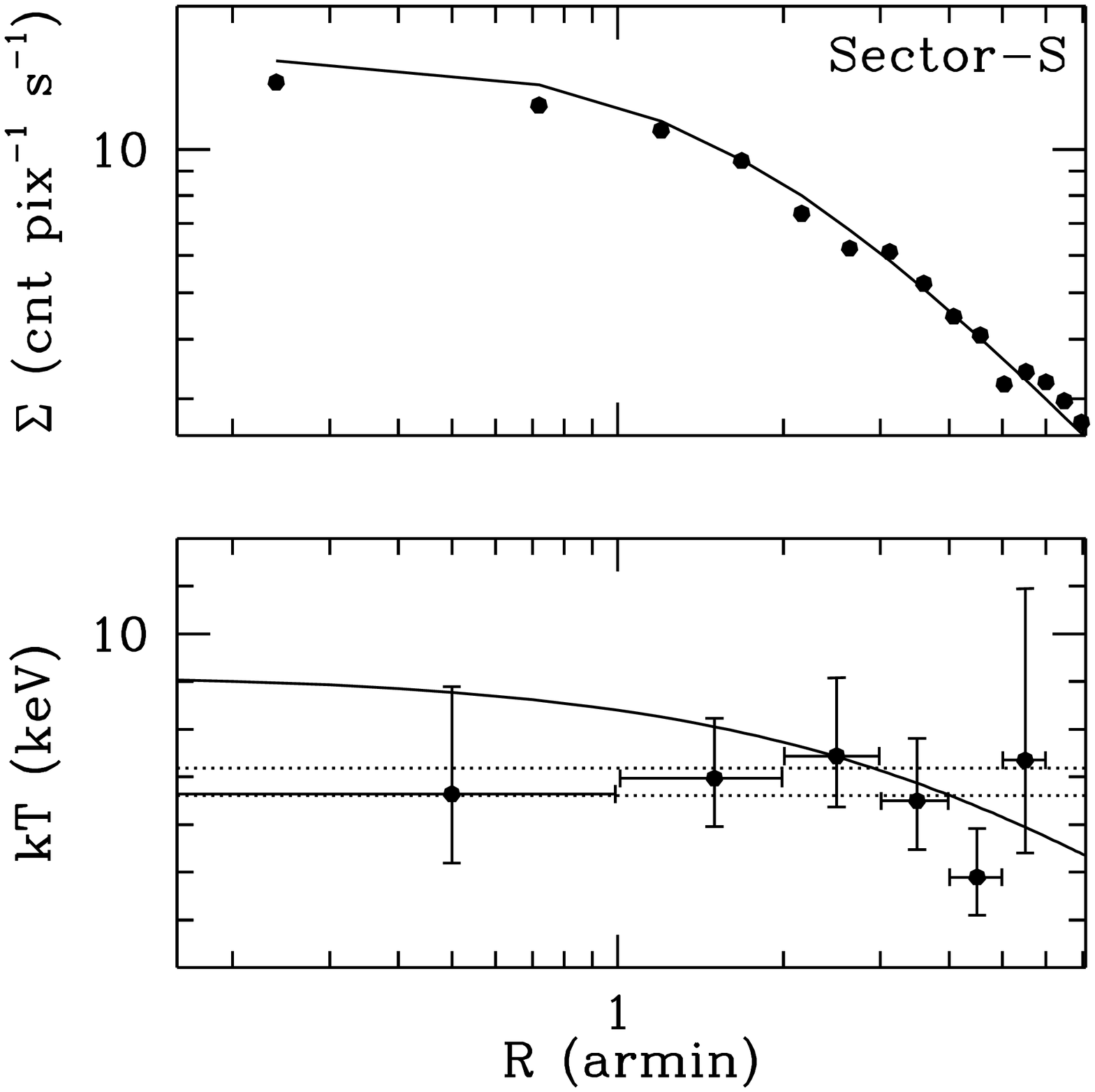}
\leavevmode 
\epsfxsize 0.48\hsize
\epsffile{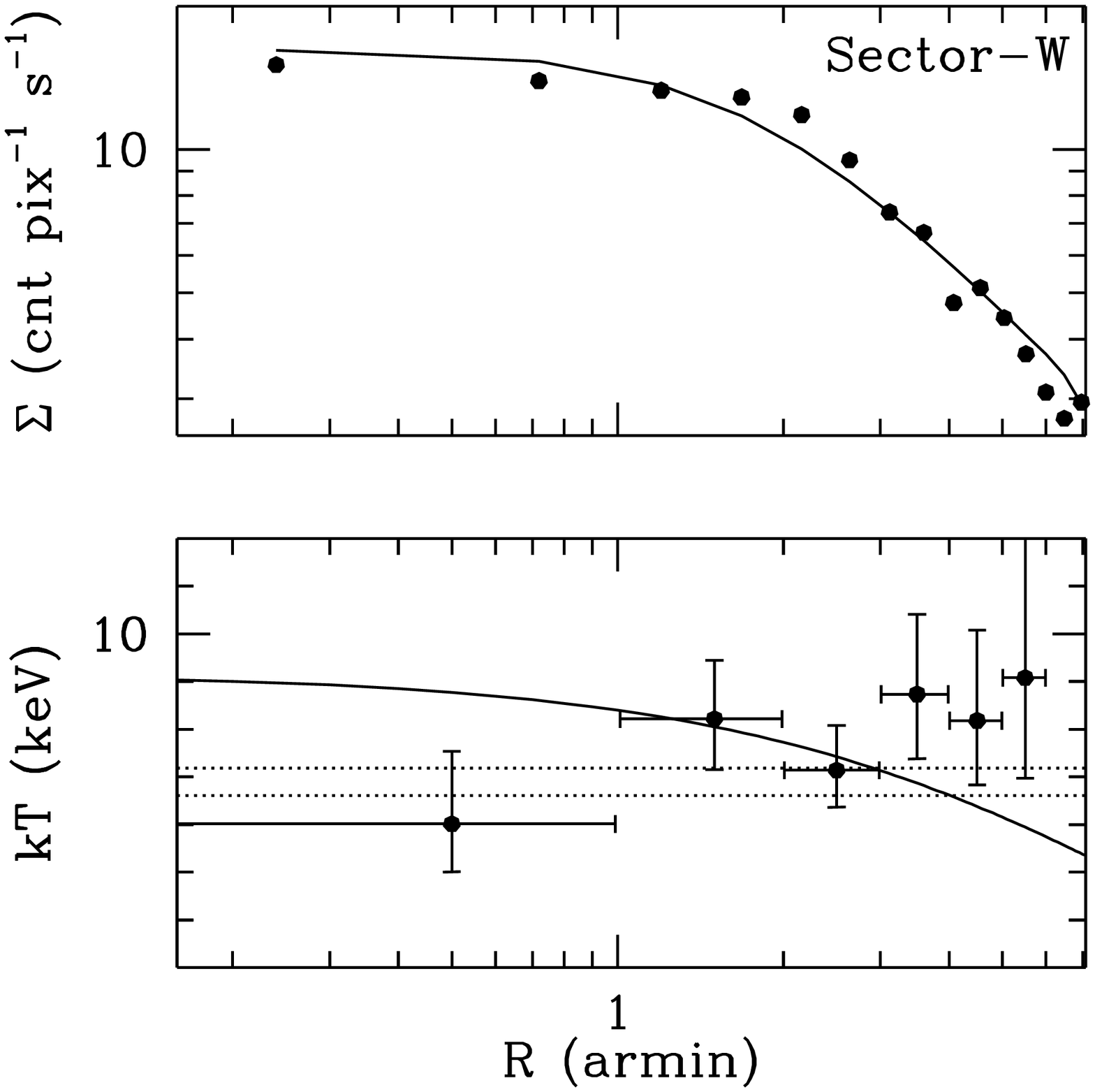}
\end{center}
\caption{Surface brightness ($\Sigma$) and temperature ($kT$) profiles
for each of the four sectors in Abell~2255 shown in
Fig.~\ref{images}(a). The surface brightness as registered on the MOS1
detector only are shown, but the profiles from the other cameras are
similar.  The solid lines in the surface brightness plots represent
the best-fitting model for each sector, found by the 2-dimensional
analysis of Section~\ref{2d}. The errors shown for the temperature are
the 90~percent errors. For the temperature profiles, data from all 3
instruments were used. The solid lines in the temperature plots
represent the universal temperature profiles of Loken et al.
(2002), and the dotted horizontal lines mark the confidence limits of
the over-all cluster temperature from Table.~2.}\label{A2255T_prof_pies}
\end{figure*}

The background-subtracted and exposure-corrected images from each \xmm\
camera were fitted in {\sc sherpa} by a 2-dimensional $\beta$-model. The
fit was restricted to the inner 10~arcmin of the image, and all bright
point sources were subtracted. The images from the three \xmm\ cameras
were fitted simultaneously. The core radius ($r_{\rm c}$), the
$\beta$-parameter, the location of the x-ray centre, the ellipticity
($e$), and position angle ($\theta$) were linked and left free, so
that their values are determined by the fitting procedure. The
normalizations of the $\beta$-models were free to vary independently
for each instrument.

This fitting procedure resulted in the following best fitting
parameters: $r_{\rm c}$ =
110.8$_{-2.3}^{+2.0}$~arcsec=165.9$_{-3.5}^{+3.0}$~kpc, $\beta$ =
0.37$\pm$0.17, $e$=0.191$\pm$0.006, $\theta$=0.048$\pm$0.015~deg. The
X-ray centre is found at $\alpha_{cen} = 17^{h}12^{m}50\fs38$
$\delta_{cen} = +64\degr03\arcmin42\farcs56$, which does not coincide
with any of the big cluster galaxies, as can be also seen in the plot
of Fig.~\ref{2D}. The values of the $\beta$-index and the $r_{\rm
c}$ derived from this fitting procedure appear lower than their
`canonical' values for clusters of galaxies. If we fix the
$\beta$-index at 0.65, the 2-dimensional analysis results in a larger
value for the $r_{\rm c}$, as expected due to the coupling between the
$\beta$-index and the $r_{\rm c}$. However, this fit does not
represent the data well, especially in the inner cluster regions. The
low $\beta$ values (with the accordingly low $r_{\rm c}$) can be
understood if one takes into account the fact that the cluster is
extended, and the \xmm\ data map only its inner regions. Its surface
brightness distribution does not reach the background levels at
$r=10$\,arcmin, which is the size of the fitting image region, and
there are some difficulties in reliably subtracting the background, as
discussed in section~2.2 above. To test for possible effects due to
residual background, we performed a 2-dimensional analysis similar to
that described above, but with a model comprised of a constant plus a
$\beta$-model. The inclusion of the constant in the model resulted in
best-fitting values for the $\beta$-index and the $r_{\rm c}$ closer
to their `canonical' values, and to the ones found by earlier
investigations. We found $\beta$-index and $r_{\rm c}$ values of
0.56$_{-0.17}^{+0.47}$ and
193.4$_{-1.9}^{+101.8}$~arcsec=289.9$_{-2.8}^{+152.6}$~kpc
respectively.

As explained in Section~2.2, we found that the blank-sky background
files for the PN instrument require large scaling factors to match the
background levels of the \cl\ observations, and we therefore decided not 
to use these blank-sky background files for spectral analysis. For
the spatial analysis, the background-subtracted images of Section~3
(which made use of the blank-sky background files)
were fitted by a 2-dimensional model. Due to the uncertainties in
the scaling factor of the PN camera, we checked the analysis,
fitting only the images from the two MOS cameras. We found
best fitting parameters consistent with those derived from
the full three datasets, as presented above.

We co-added the best fitting model images for the MOS instruments, and
a contour plot is overlaid onto the optical image of the cluster in
Fig.~\ref{2D}, where we also show the positive (white contours) and
negative (black contours) residuals of the fit. As is seen in this
image, the cluster centre found by the best-fitting model does not
coincide with any of the large galaxies in the field.  Additionally,
there are positive residuals to the East and West of the X-ray centre
that seem to coincide with with some galaxies, while the centre of the
distribution coincides with a depression in the light distribution.

\subsection{Sectors}\label{spatial_sectors}

In order to visualize better the residuals of the 2-dimensional
analysis and compare them with the temperature distribution around the
cluster centre, we present next the radial profiles along the sectors
shown in Fig.~1(a).  We obtained the radial surface brightness
profiles in the four sectors around the centre $\alpha_{cen}$,
$\delta_{cen}$ found in Section~3.1.  Each sector was 90~degrees wide
and the orientation of all four is shown in Fig.~\ref{images}(a). A
comparison of the radial profiles in each sector with the best-fitting
model for each one found by the 2-dimensional analysis in
Section~\ref{2d} is presented in Fig.~\ref{A2255T_prof_pies}. For the
surface brightness plot, only data from the MOS1 camera are shown,
although all three cameras were used for the
analysis. Figure~\ref{A2255T_prof_pies} also presents the temperature
profiles along the same sectors. Their derivation will be presented in
a later section.

\section{Spectral Analysis}\label{spectral}

For the following spectral analysis, we used the clean and filtered
event lists produced in Section~2.1.  Responses and auxiliary files
were generated with {\sc rmfgen-1.48.5} and {\sc arfgen-1.54.7}
respectively. Generally, we model in {\sc xspec} the (0.3-8.0)~keV
energy range by absorbed thermal models. 

\subsection{Over-all temperature}\label{spec_all}

\begin{figure}\label{spectrum}
\begin{center}
\setlength{\unitlength}{1cm}
\begin{picture}(8,7)
\put(-0.5,8){\includegraphics{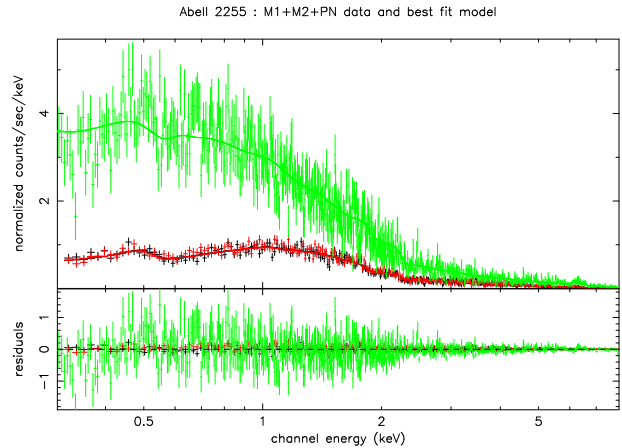}}
\end{picture}
\end{center}
\caption{\xmm\ spectra of the inner 6~acmin of Abell~2255.  The data
from the two MOS cameras are shown as black and grey crosses, and give
the same count rate. The spectra accumulated from the PN instrument
are always the ones at higher ${\rm cnt \ s^{-1} \ keV^{-1}}$. For
clarity, data from observation 0112260801 are only shown. The
best-fitting model, and the residuals of the fit are also included.}
\end{figure}

Source spectra were extracted in a circular region centred at
$\alpha_{cen}$, $\delta_{cen}$ (see Section~3.1), and extending out to
6~arcmin. The background was taken from an annular region adjacent to
the source region, between 6 and 10~arcmin from the cluster
centre. Spectra from the three \xmm\ instruments and the two
observations were fitted simultaneously by a {\it mekal} model
modified by the absorbing column ($N_{\rm H}$). During the fitting
procedure, the $N_{\rm H}$, the temperature of the plasma ($kT$), the
metallicity ($Z$), and the normalization were left free to vary.

The \xmm\ spectra, the best-fitting model, and the residuals of the fit
are shown in Fig.~\ref{spectrum}. For clarity, data only from the
observation 0112260801 are included, but the spectra accumulated from
the other observation are very similar. The results of this fitting
procedure are shown in Table~\ref{spectra_info}. In the same table we
also list the overdensity radius $R_{200}$ and the mass $M_{500}$,
derived from the cluster temperature, using the relation
of Evrard, Metzer, Navarro (1996). In
the same table we also list the value of $\beta_{\rm spec}=(\mu m_{p}
\sigma^2)/(k_{B}T_{\rm ICM})$, where $\mu$, $\sigma$ are the mean
molecular weight ($\mu=0.6$), and velocity dispersion of the cluster
($\sigma$=1200~${\rm km \ s^{-1}}$ -- see in the introduction).

\begin{table}
\caption{Spectral properties}\label{spectra_info}
\begin{center}
 \begin{tabular}{lc}   \hline \hline

Parameter		&
Value			\\

\hline

$kT$~(keV)		&
$6.90\pm0.29$		\\

$N_{\rm H}$~($\times 10^{20}\ {\rm cm^{2}}$) 		&
$1.43\pm0.29$				\\

$Z$~(Z$_{\sun}$)					&
$0.24\pm0.05$				\\

$L_{\rm x}^{1}$~($\times 10^{44}\ {\rm erg \ s^{-1}}$)			&
$2.780\pm0.042$		\\

$\chi^{2}$/d.o.f.					&
2554/2372						\\

\hline

$R_{200}$~(Mpc/arcmin)			&
2.11/23.5				\\

$M_{500}$~($\times 10^{14} \ M_{\sun}$)	&
4.44					\\

$\beta_{\rm spec}$			&
1.305					\\

\hline

\end{tabular}
\vspace{0.2cm}
\begin{minipage}{10cm}
\small NOTES: $^{1}$: (0.3-8.0)~keV, unabsorbed X-ray luminosity 

\end{minipage}
\end{center}
\end{table}

The derived cluster temperature and abundances are generally in good
agreement with the results from the {\it Einstein} (David et al. 1993;
White, Jones, \& Forman 1997), and {\it ASCA} (White 2000) satellites. A
significant discrepancy arises when comparing with the \rosat\ results
of Burns at el. (1995), and Feretti et al. (1997). This issue will be
investigated and discussed later in this paper (Section~5).

\subsection{Temperature distribution}\label{Stdist}

As mentioned in the introduction, Davis \& White (1998) showed that
the temperature distribution in \cl\ might be more complex than can be
described with just a single global temperature. Feretti et al. (1997)
also noted a difference in the hardness ratio data between the eastern
and western regions of the cluster. As will be clear from the next
sections, the \xmm\ data also argue for the existence of such an
asymmetry.

\subsubsection{Sectors}\label{spectral_sectors}

In order to disclose the azimuthal variations of the temperature
profiles, we obtained spectra in concentric annuli in the four
sectors shown in Fig.~\ref{images}(a). The width of each
annulus was 1~arcmin, yielding spectra with adequate number of counts 
for the full spectral modelling. We performed exactly the same fitting
procedures as in Section~4.2.1. The derived temperature profiles are
shown in Fig.~\ref{A2255T_prof_pies} along with the corresponding
surface brightness distributions, whose derivation was discussed
earlier in this paper.

Recently, Loken et al. (2002) derived a `universal temperature
profile' for clusters using numerically simulated clusters.  To
derive the profile, they used only simulated clusters that appeared
relaxed, discarding the ones that showed signs of recent disturbances
due to mergers.  They found that the temperature declines with the
distance ($r$) from the cluster centre as $1.33 T_{0} (1 +
1.5r/\alpha_x)^{- \delta}$~keV, where $T_{0}$ is the `global'
temperature, and $\alpha_x = r_{\rm vir}$ the virial radius of the
cluster. Fits to their simulated data led to a value for the exponent
of $\delta$ of 1.6, and the normalization of 1.33.  This
theoretical temperature profile shows the expected behaviour of a
relaxed cluster at a temperature $T_{0}$. Deviations from it should be
signs that the cluster is not relaxed.  This theoretical profile is
in good agreement with much observed data [see, for example Sakelliou
\& Ponman (2004), and references therein]. However, Loken et
al. (2002) found that they could not reproduce the central core seen
in the observational temperature profiles of De Grandi \& Molendi
(2002), and that the model over-predicts the temperature in the
central regions (for $r<0.1 r_{\rm vir}$). A disagreement between the
\xmm\ data of \cl\ and the `universal temperature profile' might also
be seen in Fig.~\ref{A2255T_prof_pies} in the inner
$r<2$~arcmin. However, it has been found that the `universal
temperature profile' of Loken et al. (2002) is in good agreement with
the observational temperature profiles at large radii. Support for this
belief comes from the recent temperature profiles obtained from
\chandra\ (Vikhlinin et al. 2005) and \xmm\ data (Piffaretti et
al. 2005), that follow very closely the theoretical profile, and argue
for a temperature decline at large radii. In the temperature plots of
Fig.~\ref{A2255T_prof_pies} we show with solid lines the temperature
profile for Abell~2255, as predicted by the above equation of Loken et
al. (2002). As $r_{\rm vir}$ we used the $R_{200}$=2.11~Mpc, and the
`global' temperature is $T_{0}$=6.90~keV given in Table~2. We have to
note though, that the profiles of Fig.~\ref{A2255T_prof_pies} do not a
extend out to large radii, but are restricted to $r<0.25R_{200}$. A
small temperature decline at $0.25R_{200}$ is expected by the
theorical profile as can be seen in Fig.~\ref{A2255T_prof_pies}.

A few striking properties for the temperature structure of \cl\ emerge
from the inspection of these figures: i) the low temperature in the
inner 2~arcmin in sector-E, and ii) the high temperature in the outer
(3-6)~arcmin region in sector-W. Along sector-N and sector-S, the
temperature profiles appear the most regular of the four, showing the
gas to be nearly isothermal at the global temperature, and consistent
with the theoretical universal temperature profile.

\begin{figure}
\begin{center} 
\leavevmode 
\epsfxsize 1.0\hsize
\epsffile{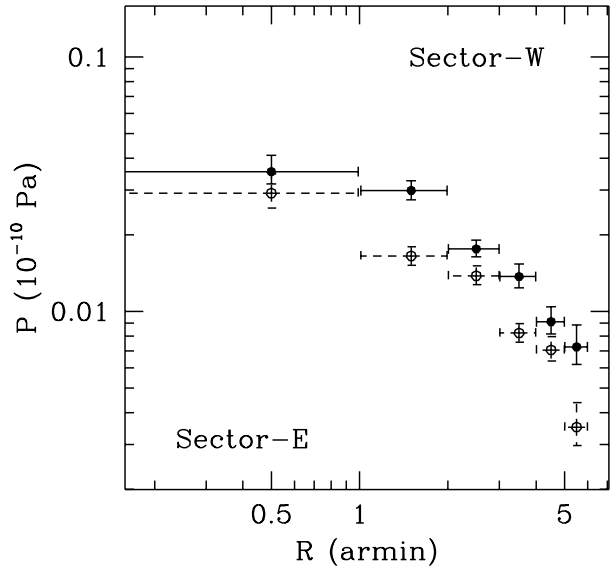}
\caption{Comparison of the pressure profiles along the sectors -W
  (filled symbols) and -E (open symbols).}\label{P}
\end{center} 
\end{figure}

The above spectral fits resulted in a normalization ($norm$) for the
{\it mekal} component in each spatial bin. Assuming that the hot X-ray
emitting component at the derived temperature is distributed uniformly
in the entire volume of each spherical bin, and using the dependency
of $norm$ on the density ($n$) ($norm \propto \int n^{2} \ dV$) we
derived a density for the plasma in each bin. With the densities and
temperatures of each bin in each sector we calculated the pressure
profiles shown in Fig.~\ref{P} where we compare the
pressures along sector-E and sector-W. The pressures for the other two
sectors (sector-N and -S) are always between the ones along sector-E
and -W. This comparison provides further evidence that the main
disturbances occur along the East-West direction of the cluster.

\subsubsection{Temperature map}\label{Stmap}

\begin{figure}\label{Tmap}
\begin{center}
\setlength{\unitlength}{1cm}
\begin{picture}(8,9)
\put(-0.5,-0.5){\includegraphics{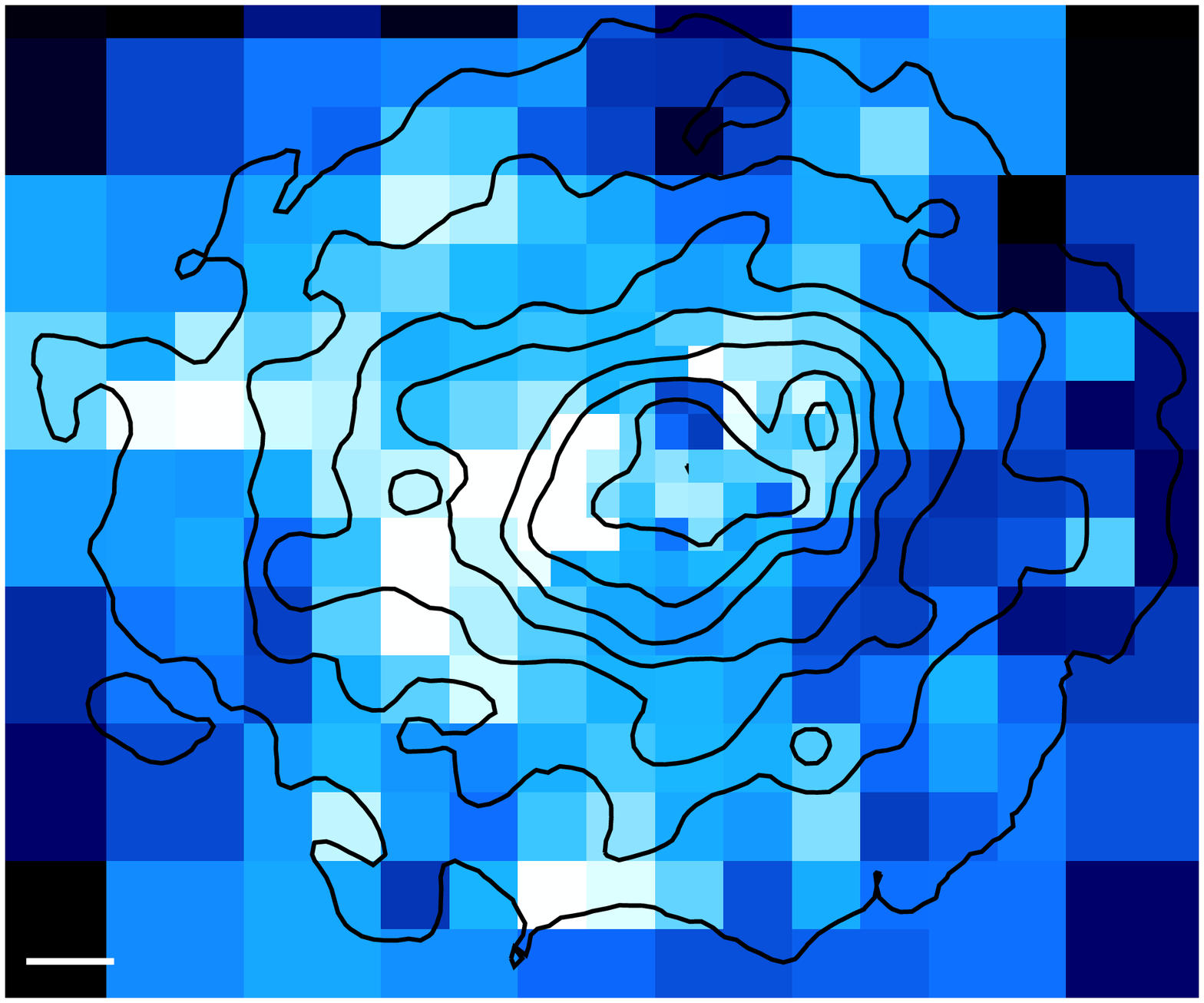}}
\put(-1.3,5.5){\includegraphics{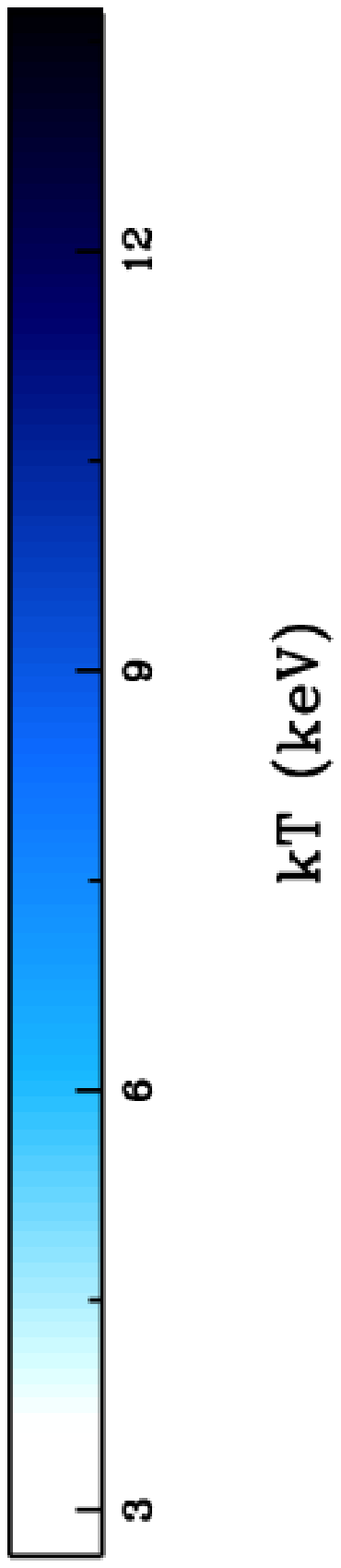}}
\end{picture}
\end{center}
\caption{The temperature map of the central region of \cl\, overlayed
by the X-ray contours shown in Fig.~\ref{images}(b). 
The horizontal bar is 1~arcmin in
length.}
\end{figure}

The analysis of the previous sections revealed an asymmetry in the
temperature distribution between the eastern and western regions of
the cluster. In order to visualize better these anomalies and
understand their 2-dimensional extent we constructed the temperature map
shown in Fig.~\ref{Tmap} by accumulating source counts in square
regions around the cluster centre. The initial region was $25 \times
25$~arcmin wide. This region was subdivided into $2 \times 2$
bins. Subsequently, each bin was divided again into smaller bins,
until the number of counts in each bin dropped below a minimum number
of counts of $n_{\rm min}=$2000, which is the minimum number of counts
we require in each bin, in order to obtain accurate temperature
values. Any bins containing $< n_{\rm min}$ adopted the properties of
the corresponding larger bin.  For the creation of the temperature map
we used only the longest of the two \xmm\ observations
(obs=0112260801). As the background we used the blank-sky background
files, and the scaling factors we found in Section~2.2. The spectral
fits for each bin were performed again in {\sc xspec}, with the
$N_{\rm H}$ being fixed to the best-fitting value found in
Section~4.1. The metal abundances were left free to vary.

To guide the eye in Fig.~\ref{Tmap} we overlay the resultant
temperature map with contours of the X-ray emission shown in
Fig.~\ref{images}(a) and (b). This image shows clearly the temperature
distribution within the core of Abell~2255, and supports the findings
of Section~4.2.2. In particular it shows that: i) the cool emission to
the East of the cluster centre is spread over a wide region, mainly
along sector-E, ii) galaxy A [see Fig.~\ref{images}(b)] is not
associated with any cold emission as will be discussed in Section~5,
but instead there is cold emission to the north-east of it, and iii)
the regions to the south and south-west of galaxy A are hotter,
reaching in some places temperatures of the order of
$\sim$(9-10)~keV. This emission does not appear associated with
  galaxy B, which is not listed as an active galaxy. Additionally,
  Fig.~1(b) shows clearly that galaxy\,B does not contribute to the
  X-ray emission.

\section{Discrepancy between \xmm\ and \rosat\ temperatures}

As noted in Section~\ref{spec_all}, a significant discrepancy arises
when comparing the temperatures derived from \xmm\ and \rosat\ : with the
\xmm\ data we find a global temperature of $\sim$6.9~keV; the spectral
fits to the \rosat\ All Sky Survey (RASS) data resulted in a much lower
temperature of $1.9_{-0.4}^{+2.3}$~keV (Burns at el. 1995) for the
central parts of Abell~2255. Feretti et al. (1997),
using pointing \rosat\ observations, derived a similarly low
temperature of $3.5 \pm 1.5$~keV. The temperature map of Davis \&
White (1998) supports the above \rosat\ findings, since they found a
temperature of $3.5_{-1.4}^{+3.7}$~keV for the inner 1.5~arcmin
region. However, it has to be noted, that none of these investigations
argue for the presence of a traditional `cooling flow' in the
cluster. Its absence is also supported by the lack of a strongly
centrally peaked surface brightness distribution (see, for example,
the surface brightness plots of Fig.~\ref{A2255T_prof_pies}).

The reason for the above discrepancy might be simply that the source
regions used for the \rosat\ analysis were not centred on the cluster
centre as defined here, but more towards the galaxies A and B [see
Fig.~\ref{images}(b)], which might host cooler X-ray atmospheres. The
recent investigation of the X-ray properties of the cluster galaxies
with \chandra\ did not include galaxy A and B (Davis et al 2003),
because unfortunately they lie on a CCD gap. As is apparent from
Fig.~\ref{images}(b) and \ref{2D}, significant emission from galaxy A
was registered by \xmm\.  Attempts to fit the \xmm\ spectra from small
regions around galaxy A did not result to any temperatures lower than
$\sim$5.6~keV. To obtain the temperature around galaxy A, we
accumulated counts in a circular region centered on the galaxy with a
0.5~arcmin radius. We fitted the (0.3-5.0)~keV spectrum with an
absorbed {\it mekal} model. The $N_{\rm H}$ and $Z$ were fixed to the
values shown in Table~2. The fitting procedure resulted to a
best-fitting temperature of 7.25$^{+2.95}_{-1.69}$~keV
($\chi^2/d.o.f.$=85/93). The temperature map of Fig.~\ref{Tmap} also
supports these findings. 

The lack of centrally concentrated cold gas is also apparent from the
analysis of the Sections~\ref{Stmap} and \ref{Stdist}. In those
Sections we found evidence for lower temperatures in the cluster
(towards its eastern regions), but again not as low as required to
support the \rosat\ results.

On the other hand, the discrepancy might be due to an inherent
artefact of the modelling of the \rosat\ data. The fits of a two (or
more) thermal components by a single one in the \rosat\ narrow energy
band might have unavoidably resulted to a single cluster temperature
that is lower than it should. As we found in the previous sections the
inner regions of \cl\ are dominated by two temperatures of $\sim$5.5
and $\sim$7~keV. Recently, Mazzotta et al. (2004) investigated the
effect of a single temperature model fits to a two temperature plasma
when observed with \chandra\ or \xmm\. They found that in some cases
the single temperature fits might be `acceptable' (regarding the
reduced $\chi^2$ of the fit), leading to a temperature that does not
correspond to any real temperature in the cluster.

In order to investigate the above possibility as the explanation for
the discrepancy between \rosat\ and \xmm\, we simulated in {\sc xspec}
\rosat\ PSPC spectra comprised of two thermal models: one at a
temperature of 5.5~keV and the second at 7~keV.  The metallicity of
both was equal to the metallicity we derived in Section~\ref{spec_all}
($Z$=0.24), and both models were absorbed by the Galactic column. We
also assumed an equal contribution by both to the composite thermal
model, choosing equal normalizations of the two models, that
additionally yielded a luminosity equal to the $L_{\rm x}$ of Table~2.
The choice for equal contributions from the two temperatures is
justified by the fact that the normalizations we find from the
spectral fits shown in Fig.~\ref{A2255T_prof_pies} for the low and
high temperatures are very similar.  The exposure time of the simulated
spectrum was set to that of the \rosat\ PSPC observation
($\sim$15~ksec). 

We modelled in {\sc XSPEC} the (0.1-2.4)~keV simulated spectrum by a
single {\it Raymond-Smith} model, using the \rosat\ response
pspcb\_gain1\_34.rsp.  The fitting procedure found a temperature of
7.6$_{-1.8}^{+3.6}$~keV and reasonable values for the $N_{\rm H}$ and
$Z$. However, the quality of the fit was poor, resulting to a
$\chi^2_{\nu} \sim 0.5$. This result indicates that the reason for the
low \rosat\ temperatures is not due to the incorrect modelling of the
\rosat\ spectrum. We have to note though, that the \rosat\ PSPC
spectrum, the quality and residuals of the fit that resulted to the
previously quoted low temperatures have not been presented, making a
further comparison with our results difficult.  As mentioned before,
the \xmm\ temperatures are consistent with the ones found by {\it
Einstein} and {\it ASCA}. Unfortunately, the disagreement with the
\rosat\ results still remains, but further investigations are outside
the scope of this paper.

\section{Summary and Discussion}

The \xmm\ data and analysis, presented in the previous sections show
strong evidence that \cl\ is far from our idealized picture of a
`relaxed' cluster. It has suffered a merger event in its recent past,
and the signatures of such a turmoil are still visible: the X-ray
emission is elongated along the East-West direction, aligned with a
chain of galaxies; the temperature distribution does not follow the
universal temperature profile of Loken et al. (2002); the X-ray
emission is not centered at any galaxy.  Especially, the temperature
structure shows the kind of disturbances expected during/after a
merger event. Although we derive a `global' temperature for the
cluster of $\sim$6.9~keV, it is not isothermal at this
temperature. Its eastern regions are cooler at $\sim$(5-6)~keV, and
towards the West the temperature reaches $\sim$8.5~keV. Following the
temperature asymmetry, there is a pressure imbalance between East and
West (see Fig.~\ref{P}), which is mainly driven by the temperature
inequality.

Over the last years, thanks to the \xmm\ and \chandra\, we have
witnessed the complicated structure of the ICM in merging/evolving
clusters. Additionally, numerical simulations of merging clusters
have advanced to such an extent that can show us the details of the
evolution of a merger event. The challenge now is to match the
observations with the simulations and understand how the final cluster
is formed.

For unequal mass mergers, simulations show (e.g., Takizawa 1999,
Roettiger et al. 1997) that after the cores of the two subclusters
collide, the smaller and cooler cluster continues travelling way from
the collision location, leaving a cool trail along its trajectory. Two
shocks are generated during the collision, that travel along the
collision axis towards opposite directions. One `front' shock is
leading the smaller subcluster, and it is diffuse. The `back' one is
more compact, and propagates towards the opposite direction.  The
X-ray properties of \cl\ support the idea 
that it is a merger remnant after the
core collision phase.  Specifically, there is a 
correspondence with the results, for example, presented by Takizawa
(1999). A comparison of Fig.~\ref{Tmap} with the temperature structure
in the merger remnant of his fig.~8 at $t=4.75$~Gyr shows a striking
similarity.  This correspondence can be also found with other
numerical work.

The comparison of the temperature distributions along sector-E and -W
of Fig.~\ref{A2255T_prof_pies} shows that there is a temperature
increase from the cluster average ($\sim$6.9~keV) up to
$\sim$8.5~keV. A possible explanation for this is that it is due to a
shock wave.  If this is the case, using eqn.~(2) from Markevitch,
Sarazin \& Vikhlinin (1999), we find that the Mach number of the
relative motion is $M \sim 1.24$, which implies a velocity of $\simeq
2400 \ {\rm km \ s^{-1}}$ (the sound speed in Abell~2255 is $c_s=1940\
{\rm km \ s^{-1}}$), and a compression factor of 1/$x \sim 1.36$. For the
above calculation we used $\gamma = 5/3$, a preshock temperature of
k$T_0\sim6.9$~keV, and postshock temperature of k$T_1\sim8.5$~keV.  If
this is the `back' shock seen in the simulations, and it has been
travelling at $v=2400 \ {\rm km \ s^{-1}}$, and if the core crossing
happened where the current cluster centre is, we find that the cores
collided some $t = s/v \sim (4~{\rm arcmin}) / (2400 \ {\rm km /
s^{-1}}) \sim 0.15~{\rm Gyr}$ ago.  Such a shock wave would have increased
the flux by an amount of $\Sigma_1/\Sigma_0=1.36^{2} \sim 1.85$. If we
compare the flux at a distance of $\sim5$~arcmin from the cluster
centre in Sector-W with the flux along sector-S and -N, we find that
the measured $\Sigma_1/\Sigma_0$ is not more than $\sim$2.  Of course,
the details of the structure depends on the initial conditions and
projection effects.

Being guided again by the numerical work [e.g., Takizawa (1999),
Randall, Sarazin, Ricker (2002)] we note that the collision of the
subclusters' cores during the `core-crossing' phase results in a sharp
and brief increase of the luminosity and temperature. Afterwards, the
cluster expands adiabatically, resulting in a decrease of its
temperature and luminosity. The luminosity drops below its initial
value, which is defined as the sum of the initial luminosities of the
two subunits. The temperature on the other hand, remains at the same
levels as the initial temperature (before the collision) until a later
stage of the merging process, at which it increases more mildly, due
to the collapse and accumulation of the cluster material towards the
new cluster centre. Thus, if a cluster is at a stage after the
dramatic `core-crossing' phase, its luminosity should be high, but not
as high as during the violent sub-clusters collision, and lower than
the initial total luminosity of the system.

In order to locate \cl\ on the L-T relation, we compare its properties
with the L-T relation derived by Markevitch (1998). We calculated the
cluster bolometric luminosity, and extrapolated it out to
1$h^{-1}$~Mpc, in the same manner as in Markevitch (1998). Although
\cl\ does not host a traditional `cooling flow', we exclude the inner
regions of the cluster to be consistent with the analysis of
Markevitch (1998). The derived luminosity, $L_{\rm bol} \simeq 12.13
\times 10^{44} {\rm erg \ s^{-1}}$, is almost double the expected
$L_{\rm bol}$ of a $\sim$7~keV cluster, according to the L-T relation,
if we compare with fig.~2 in Markevitch (1998).  During the next
stages of the merging process, the luminosity of the remnant will
remain almost unaltered, while its temperature will increase as
explained earlier. This temperature increase may be such that its
temperature and luminosity come into agreement with the L-T relation,
and will stay at that condition until the final merger remnant is
formed.  Being guided again by the the numerical simulations we find
that \cl\ will settle down to a single remnant in some $\sim$(2-3)~Gyr.

\section*{Acknowledgments}

The Digitized Sky Survey (DSS), and the NASA/IPAC Extragalactic
Database (NED) have been used. The present work is based on
observations obtained with {\it XMM-Newton}, an ESA science mission with
instruments and contributions directly funded by ESA Member States and
the USA (NASA). IS acknowledges the support of the European Community
under a Marie Curie Intra-European Fellowship.


\begin{thebibliography}{99}

\bibitem[\protect\citeauthoryear{}{}]{} Buote D.A.,  2001, ApJ, 553, L15


\bibitem[\protect\citeauthoryear{}{}]{} Burns J.O., Roettinger K.,
Ledlow M., Klypin A., 1994, ApJL, 427, L87

\bibitem[\protect\citeauthoryear{}{}]{} Burns J.O., Roettinger K.,
Pinkney J., Perley R.A., Owen F.N., Voges W., 1995, ApJ, 446, 583

\bibitem[\protect\citeauthoryear{}{}]{} David L.P., Slyz A., Jones C.,
Forman W., Vrtilek S.D., Arnaud K.A., 1993, ApJ, 412, 479

\bibitem[\protect\citeauthoryear{}{}]{} Davis D.S., White III R.E.,
1998, ApJ, 492, 57

\bibitem[\protect\citeauthoryear{}{}]{} Davis D.S., Miller N.A.,
Mushotzky R.F., 2003, ApJ, 597, 202 

\bibitem[\protect\citeauthoryear{}{}]{} De Grandi S., Modendi S., 2002
  ApJ, 567, 163


\bibitem[\protect\citeauthoryear{}{}]{} Evrard A.E., Metzler C.A.,
  Navarro J.F., 1996, ApJ, 469, 494

\bibitem[\protect\citeauthoryear{}{}]{} Feretti L., B\"{o}hringer H.,
Giovannini G., Neumann D., 1997, A\&A, 317, 432

\bibitem[\protect\citeauthoryear{}{}]{} Giovannini G., Feretti L.,
2000, New Astronomy, 5, 335


\bibitem[\protect\citeauthoryear{}{}]{} Giovannini G., Tordi M.,
Feretti L., 1999, New Astronomy, 4, 141

\bibitem[\protect\citeauthoryear{}{}]{} Govoni F., Murgia M., Feretti
  L., Giovannini G., Dallacasa D., Taylor G.B., 2005, A\&A, 430, L5 

\bibitem[\protect\citeauthoryear{}{}]{} Loken C., Norman M.L., Nelson
  E., Burns J.O., Bryan G.L., Motl P., 2002, ApJ, 579, 571   

\bibitem[\protect\citeauthoryear{Lumb}{2002}]{l} Lumb D. 2002, `{\it
EPIC BACKGROUND FILES}, XMM-SOC-CAL-TN-0016, issue 2.0

\bibitem[\protect\citeauthoryear{}{}]{} Markevitch M., 
1998, ApJ, 504, 27

\bibitem[\protect\citeauthoryear{}{}]{} Markevitch M., Sarazin C.L.,
Vikhlinin A., 1999, ApJ, 521, 526

\bibitem[\protect\citeauthoryear{}{}]{} Mazzotta P., Rasia E.,
  Moscardini L., Tormen G., 2004, MNRAS, 354, 10

\bibitem[\protect\citeauthoryear{}{}]{} Mullis C.R., Henry J.P., Gioia
I.M., B\"{o}hringer H., Briel U.G., Voges W., Huchra J.P., 2001, ApJL,
553, L115

\bibitem[\protect\citeauthoryear{}{}]{} Piffaretti R., Jetzer Ph.,
Kaastra J.S., Tamura T., 2005, A\&A, 433, 101

\bibitem[\protect\citeauthoryear{}{}]{} Pratt G.W., Arnaud M., 2002, A\&A, 394, 375


\bibitem[\protect\citeauthoryear{}{}]{} Randall S.W., Sarazin C.,
  Ricker P.M., 2002, ApJ, 577, 579

\bibitem[\protect\citeauthoryear{}{}]{} Roettiger K., Loken C., Burns
  J.O., 1997, ApJS, 109, 307 

\bibitem[\protect\citeauthoryear{}{}]{} Sakelliou I., Ponman T.J.,
  2004, MNRAS, 351, 1439

\bibitem[\protect\citeauthoryear{}{}]{} Takizawa M., 1999, ApJ, 520, 514

\bibitem[\protect\citeauthoryear{}{}]{} Vikhlinin A., Markevitch M.,
  Murray S.S., Jones C., Forman W., Van Speybroeck L., 2005, astro-ph/0412306

\bibitem[\protect\citeauthoryear{}{}]{} White D.A., 2000, MNRAS, 312, 663

\bibitem[\protect\citeauthoryear{}{}]{} White D.A., Jones
C., Forman W., 1997, MNRAS, 292, 419

\bibitem[\protect\citeauthoryear{}{}]{} Yuan Q., Zhou X., Jiang Z., 2003, ApJS, 149, 53

\end{thebibliography}
\end{document}